\documentstyle[prl,twocolumn,aps]{revtex}
\input{epsf}

\newcommand{\bq}{\begin{equation}}
\newcommand{\ee}{\end{equation}}
\newcommand{\fr}[2]{\frac{#1}{#2}}
\newcommand{\eps}{\varepsilon}

\begin{document}
\draft

\title{Chaos Thresholds in finite Fermi systems}

\author{P.G.Silvestrov}

\address{Budker Institute of Nuclear Physics, 630090
Novosibirsk, Russia}

\maketitle

\begin{abstract}

The development of Quantum Chaos in finite interacting Fermi
systems is considered. At sufficiently high excitation energy
the direct two-particle interaction may mix into an eigen-state
the exponentially large number of simple Slater-determinant
states.  Nevertheless, the transition from Poisson to
Wigner-Dyson statistics of energy levels is governed by the
effective high order interaction between states very distant in
the Fock space.  The concrete form of the transition depends on
the way one chooses to work out the problem of factorial
divergency of the number of Feynman diagrams. In the proposed
scheme the change of statistics has a form of narrow phase
transition and may happen even below the direct interaction
threshold.

\end{abstract}


\section{Introduction}\label{sec:1}

The investigation of chaotic properties of finite interacting
quantum systems has now a few decades history
\cite{Wigner,Wong,Bohigas,Zelev,Flamb,Pichard,Shepelyansky}. The
Random Matrix Theory, originally introduced for modelling this
problem, adequately describes the statistics of complicated high
energy excitations (see Ref.~\cite{Weidenmuller} for an
extensive review). However, the concrete mechanism of developing
the chaotic behavior in the finite Fermi system with increasing
of excitation energy still requires the further understanding.
The interest in this problem has even increased in last years in
particular due to the recent explosion of activity in the field
of Mesoscopic Physics (quantum dots, atomic clusters, helium
droplets etc.).  The experimental observation of electronic
excitations in quantum dots \cite{SImry} has stimulated the
progress \cite{Sivan,Blanter,AGKL,Mirlin,me} in theoretical
investigation of chaotic disintegration of single electron
excitations in this systems. The single particle excitations,
being most easily produced experimentally, constitute an
important part of Hilbert space of complicated system. However,
the number of such states is exponentially small compared to the
total number of excited states. The onset of chaos in the
majority of many-particle states is the subject of this paper.

We will consider finite, but sufficiently large Fermi systems.
The particles are supposed to occupy almost all orbitals below
the Fermi energy $\eps_F$ of mean field single particle
Hamiltonian \cite{endnote}. The averaged single particle level
spacing is $\Delta \ll \eps_F$. We will denote as simple states
the single Slater determinant states having definite occupation
numbers ($0$ or $1$) of all orbitals. The simple states are
mixed by the two-particle interaction and the matrix elements
$V$ are supposed to be small and Gaussian random
\bq\label{g}
\overline{ V^2 } = \Delta^2/g^2 \ \ , \ \ g\gg 1 \ \ .
\ee
Here $g$ may be interpreted as dimensionless conductance in the
case of quantum dot \cite{Blanter,AGKL}. This simple model was
in fact introduced long ago in Refs. \cite{Wong,Bohigas}.

Due to the weak interaction Eq. (\ref{g}), the low energy
excited states are almost unperturbed. With the increase of
energy ${\cal E}$ the density of states grows rapidly and the
exact wave functions at ${\cal E}\gg \Delta$ are formed by many
chaotically mixed simple states. However, this chaotization of
excited states proceeds in a very nonuniform and complicated
way. For illustration we have shown on the Fig.~1 different
energy scales relevant to this problem.

The complexity of excited state is usually characterized by the
inverse participation ratio (IPR) $I=\sum_k|\psi_i(k)|^4$, where
$\psi_i(k)$ is the amplitude of $k$-th simple excitation in the
$i$-th exact wave function. Physically $I$ is the effective
inverse number of simple states mixed into exact one.

While the density of single particle states $\nu_1=1/\Delta$
does not depend on energy, the total density grows extremely
fast with ${\cal E}$ being $\nu_{total}\sim
\Delta^{-1}\exp(2\pi\sqrt{{\cal E}/6\Delta})$ \cite{Bohr}. We
use large ${\cal E}$ for the total energy of the many-particle
configuration and small $\eps$ for the energy of single-particle
orbitals. Due to the huge number of states (analog of the
micro-canonical ensemble in thermodynamics) the averaged
occupation number for given orbital for arbitrarily chosen
"typical" simple state is given by the usual Fermi-Dirac
distribution
\bq\label{FD}
f(\eps-\eps_F)=
\overline{n(\eps)}=\left(\exp
\left[\fr{\eps-\eps_F}{T}\right] +1\right)^{-1} \ \ ,
\ee
where $ T=\sqrt{6 {\cal E}\Delta}/\pi$ \cite{BlBl}. Note, that
we still have not taken into account the interaction $V$, but
only explore the wide statistics of single-Slater-determinant
states.

The low energy thresholds on the Fig.~1 correspond to the decay
of single quasi-particle. Quasi-particles in quantum dot may be
seen as peaks broadened by the interaction (bunches of
$\delta$-peaks) in the one particle spectral density up to the
energy ${\eps} \sim \Delta g$ \cite{Sivan}. Above ${\eps} \sim
\Delta\sqrt{g}$ the broadening arise due to a decay of
quasi-particle into two particles and one hole and may be
explained by the Fermi golden rule. Below this energy the width
predicted by golden rule turns out to be smaller than
three-particle level spacing.  However, as it was shown recently
\cite{AGKL} up to energy ${\eps}\sim \Delta\sqrt{g/\ln g}$ the
quasi-particle disintegration still proceeds through the
effective interaction with the many excited particle and hole
states. A nonuniform distribution of the matrix elements of
effective interaction in this case leads to a very unusual
distribution of spacings between peaks and widths of the
bunches\cite{me}.

The original intention of this research was to apply the methods
developed for description of quasi-particle disintegration for
the investigation of the full spectrum of complicated system.
However, the onset of chaos in the many-particle excitations
provided us with a number of surprising new effects not present
in the single quasi-particle case. Only the estimate of the
effective high order matrix element in the section 3 may be
considered as straightforward generalization of the analogous
estimate in the Ref.\cite{me}. In the section 2 we will show,
how the direct interaction between particles may mix into one
wave function an exponentially large number of simple states
even inside the region of Poissonian spectrum. In the section 3
we consider the self-consistent scheme of onset of chaos based
on the analysis of the scaling behavior of the direct and
effective interaction.  However, in the following section 4 we
propose an another scenario of the transition to Wigner-Dyson
caused by the factorial divergency of the high order
interaction. The detailed comparison of the features of these
two scenarios is given at the end of paper (Sec.~5).

\section{Mixing by the direct interaction}\label{sec:2}

For weak interaction the strong mixing of different quantum
states may take place only if their energies occasionally turn
out to be very close. Therefore, consider first only two states
$|i\rangle$ and $|f\rangle$ with energy difference $\eps$
connected by matrix element $V$. The simple calculation
\cite{Ponomarev} gives the IPR for exact wave function $\alpha
|i\rangle+\beta |f\rangle$
\bq\label{pr}
I = \alpha^4 +\beta^4 = 1 -
\fr{2V^2}{\eps^2 + 4V^2} \ \ .
\ee
Let us fix the state $|i\rangle$ and consider a set of states
$|f\rangle$ with a density $dn/d\eps=\Delta_f^{-1}$.  It is
convenient to introduce the intermediate scale $\eps_0$; $|V|\ll
\eps_0 \ll \Delta_f$. With the probability $1-2\eps_0/\Delta_f$
there are no final states falling into the strip $-\eps_0<
\eps<\eps_0$ and $I=1$ up to negligible correction $\sim V^2/
\Delta_f\eps_0$. With the small probability $2\eps_0/\Delta_f$ one of
the states $|f\rangle$ enters the strip. In order to find $I$ in
this case one has to average the Eq.~(\ref{pr}) over the
interval of energy
\bq\label{Blant}
\int_{-\eps_0}^{\eps_0} \left( 1 -
\fr{2V^2}{\eps^2 + 4V^2} \right) \fr{d\eps}{2\eps_0}= 1-
 \fr{2\pi{|V|}}{2\eps_0}
\ee
and then average the $|V|$ over
the ensemble of matrix elements. Combining the two
contributions one finds
\bq\label{P}
I= 1-\fr{2\eps_0}{\Delta_f} + \fr{2\eps_0}{\Delta_f} \left( 1-
 \fr{2\pi\overline{|V|}}{2\eps_0} \right)
=1 -2\pi \fr{\overline{|V|}}{\Delta_f}  \ .
\ee
This simple formula is basic for our perturbative estimates. The
averaging of the modulus of $V$ for strongly non-Gaussian high
order effective matrix elements will be the main subject of the
next section.

The goal of this paper is to consider the mixing of complicated
excited states. Occupation numbers for these states are
described by the Eq. (\ref{FD}). The density of directly
accessible states, or the density of pairs of occupied $1,2$ and
unoccupied $3,4$ states with $\eps_1+\eps_2=\eps_3+\eps_4$ which
may be connected directly by the interaction is
\bq\label{Delta4}
\fr{1}{\Delta_2}=\fr{1}{(2!)^2}\int \delta(\sum x_i)
\prod_{i=1}^4 f_i
\fr{dx_i}{\Delta}
= \fr{T^3}{\Delta^4} \fr{\pi^2}{6} \ \ .
\ee
Here $f_i = f(x_i)$ (\ref{FD}), $x_{1,2}=\eps_{1,2}-\eps_F, \
x_{3,4}=\eps_F-\eps_{3,4}$ and we have used that $1-f(x)=f(-x)$.
The factor $(2!)^{-2}$ accounts for the identical initial and
final particles. It is easy now to use the Eq. (\ref{P}) in
order to find the IPR. However, this result may be even further
improved.  Eq. (\ref{P}) with $\Delta_f=\Delta_2$ describes the
situation when only for one two-by-two transition
$1,2\rightarrow 3,4$ the energy difference turns out to be of
the same order of magnitude with the matrix element. In the
second order in $V$ the leading contribution to $I$ is given by
double event $1,2\rightarrow 3,4$ and $5,6\rightarrow 7,8$.
However, as far as none of the orbitals $5-8$ not coincides
with any of $1-4$ these two corrections are completely
independent and corresponding contributions to $I$ should be
simply multiplied. Effectively, all one-particle states are
chosen from $\sim T/\Delta$ orbitals around $\eps_F$. Therefore,
if the number of couples of pairs falling into the energy
interval $\sim |V|$ is small compared to $\sim T/\Delta$ one may
neglect the interference between different direct transitions.
Again we may introduce the intermediate scale $\eps_0$;
$|V|\ll\eps_0$. The number of directly accessible final states
is $2\eps_0/\Delta_2\gg 1$, but until $\eps_0/\Delta_2\ll
T/\Delta$ the IPR is found simply as a product over these
channels
\begin{eqnarray}\label{inv}
I =
\left(1- 2\pi \fr{\overline{|V|}}{2\eps_0}
\right)^{\fr{2\eps_0}{\Delta_2}} &=& \exp\left\{ -2\pi
\fr{\overline{|V|}}{\Delta_2}\right\} \\
&=&\exp\left\{ -\fr{4\sqrt{3}}{\sqrt{\pi}g}
\left( \fr{{\cal E}}{\Delta}\right)^{\fr{3}{2}}\right\} \ .
\nonumber
\end{eqnarray}
Here we have averaged $|V|$ for the Gaussian random interaction
Eq.~(\ref{g}).  This result is valid for $\Delta/g\ll \Delta_2
T/\Delta$, or $T\ll \Delta\sqrt{g}$. The exponentiation of the
IPR is a direct consequence of the two-particle character of the
interaction. It would be very inconvenient to investigate such
effects if one will consider the Hamiltonian of the system as a
random (sparsed) Matrix in the full Hilbert space. Surprisingly,
to the best of my knowledge, the simple effect Eq.  (\ref{inv})
has not been considered yet e.g. for description of the
excitations in complex nuclei.

Due to the Eq.~(\ref{inv}) at ${\cal E}> \Delta g^{2/3}$ the
number of components in typical wave function becomes
exponentially large. However, within the range of validity of
the Eq.~(\ref{inv}) only a small (also exponentially) part of
close energy levels is mixed in any exact eigenstate. Therefore
this simple effect is not enough for creation of full Quantum
Chaos.

\section{High order effective interaction}\label{sec:3}

Corrections to the Eq.~(\ref{inv}) arise after taking into
account of the high order effective interaction between
three-particle states, then four-particle etc. The example of
corresponding tree-type Feynman diagram is shown on Fig.~2. The
tree-type diagrams are maximally enhanced due to a wide
statistics of initial and final orbitals. Contributions from the
diagrams with closed loops are small like some power of
$\Delta/T$.  Generalization of the Eq. (\ref{Delta4}) for the
case of $n$-particle interaction gives
\bq\label{Delt}
\fr{1}{\Delta_{n}} = \int \fr{\delta (\sum x_i)}{(n!)^2}
\prod_{i \le 2n} f_i \fr{dx_i}{\Delta} =
\fr{(2\pi T)^{2n-1}}{\pi n(2n)!\Delta^{2n}} \ .
\ee
Again $(n!)^{-2}$ accounts for the identical fermions. Note
that $(2n)!\sim 2^{2n}(n!)^2$ and therefore the integration over
the energies $x_i$ does not lead here to any $n!$ suppression.
This is in strict contrast with what happens for decay of
quasi-particle \cite{me}, where the main suppression of the
density of final states $\nu_n$ arose due to the integration
over the energies of final particles.

Sufficiently more complicated turns out to be the calculation of
averaged modulus of the effective matrix element entering both the
Eq.~(\ref{P}) and the Eq.~(\ref{inv}). The simplest variant of
the effective interaction is given by the three-particle
interaction arising in the second order of perturbation theory
\bq\label{Veff3}
{ V_{eff}^{(3)} } =
\sum_2 \fr{V_{12} V_{23}}{\eps_1 -\eps_2}
 \ .
\ee
Formally this interaction mixes together three states
$|1\rangle$, $|2\rangle$, $|3\rangle$. The use of the concept of
effective interaction requires that the admixture of the
intermediate state $|2\rangle$ in the total wave function is
small compared to those of $|1\rangle$ and $|3\rangle$. This
means that ${V_{12},V_{23}}\ll|\eps_1 -\eps_2|\approx |\eps_2
-\eps_3|$. The contribution to averaged IPR from strongly mixed
(almost degenerate) $|1\rangle$, $|2\rangle$ and $|3\rangle$
should be considered separately, but this correction is small
like $1/\ln g$ compared to the total contribution of the
effective interaction. On the other hand, we will be interested
in the tail of distribution of the effective matrix elements.
For Gaussian distributed $V$ (\ref{g}) the probability to find a
large individual matrix element of the original two-particle
interaction is exponentially small. The large matrix elements
appears if one term in the sum in Eq.~(\ref{Veff3}) occasionally
has anomalously small denominator $|\eps_1 -\eps_2|\ll \Delta$.
All the other (random) contributions in the sum in this case may
be neglected for $\ln g\gg 1$. The result of this lengthy
discussion may be
summarized in the formula for the probability distribution for
the effective matrix element (\ref{Veff3})
\bq\label{P3}
P^{(3)}(V)\sim \fr{\Delta}{g^2}\fr{1}{V^2} \ ; \
g^{-2}<|V|/\Delta<g^{-1} \ .
\ee
The averaged modulus of the matrix element in this case is
$\overline{|V_{eff}^{(3)}|}\sim \Delta \ln g/g^2$.

Higher order effective matrix elements are obtained by simple
adding of more intermediate states into the
Eq.~(\ref{Veff3}). The extensive discussions of the problems
arising while calculating the averaged value of this matrix
element may be found in Ref.~\cite{me}. For example, the
averaged modulus of the four-particle effective matrix element
for given Feynman diagram is given by
\begin{eqnarray}\label{Veff4}
\overline{\left| V_{eff}^{(4)}\right| } &=&
\overline{\left|\sum_{2,3} \fr{V_{12} V_{23} V_{34}}{
\eps_2\eps_3} \right|} =\\ \ \ \ &&
\fr{(\sqrt{2}\Delta)^3}{(\sqrt{\pi}g)^3} \int
\fr{1}{|\eps_2
\eps_3|}\fr{d\eps_2}{2\Delta}\fr{d\eps_3}{2\Delta} 
 \ . \nonumber
\end{eqnarray}
Here compared to the Eq.~(\ref{Veff3}) we have made the change
of variables $\eps_1-\eps_2 \rightarrow \eps_2$ and
$\eps_1-\eps_3 \rightarrow \eps_3$. The pre-integral factor
$\sim (\Delta/g)^3$ in the Eq.~(\ref{Veff4}) arose after
averaging of Gaussian distributed (\ref{g}) direct matrix
elements $|V_{ij}|$. The upper bound for the both remaining
integrals in the Eq.~(\ref{Veff4}) is given by the
single-particle level spacing $|\eps_{2,3}|< \Delta$. Due to the
logarithmic character of the integrals the order of magnitude
estimate of the bounds is enough. More interesting is the origin
of the bounds at small $\eps_{2,3}$. Analogously to what we have
discussed for the three-particle interaction the
formula~(\ref{Veff4}) describes the mixing of four states
$|1\rangle$, $|2\rangle$, $|3\rangle$, $|4\rangle$ in the third
order of perturbation theory. The interaction of the four states
may be replaced by the effective interaction of only two
$|1\rangle$ and $|4\rangle$ if the admixture of intermediate
states $|2\rangle$ and $|3\rangle$ is small. This requirement
for example for state $|2\rangle$ means that $|\eps_2| \gg
|V_{12}|\sim \Delta/g$ (weak mixing of $|2\rangle$ and
$|1\rangle$) and $|\eps_2|\gg |V_{23} V_{34}/\eps_3| \sim
(\Delta/g)^2/|\eps_3|$ (weak mixing of $|2\rangle$ and
$|4\rangle$). Another two inequalities for the vector
$|3\rangle$ are $|\eps_3| \gg \Delta/g$ and $|\eps_3| \gg
(\Delta/g)^2/|\eps_2|$. Combining all this inequalities one
finds for the range of integration in the Eq.~(\ref{Veff4})
\bq\label{bounds}
\Delta/g\ll |\eps_{2,3}| \ll \Delta \ .
\ee
These bounds allows one to find the averaged matrix element with
the $\sim 1/\ln g$ accuracy: $\overline{|V_{eff}^{(4)}|}\sim
\Delta (\ln g)^2/g^3$. In the similar way one may consider the
five-particle, six-particle, etc. effective interaction.
Unfortunately, the system of inequalities describing the range
of integratin for the intermediate energies turns out to be
sufficiently more complicated in these cases.

The formulas like Eq.~(\ref{Veff4}) may be used to find again
the distribution of the effective matrix elements. In particular
for the four-particle interaction
\begin{eqnarray}\label{PVeff4}
&&\overline{\left| V_{eff}^{(4)}\right| } = \int
P^{(4)}(V_{eff}^{(4)}) dV_{eff}^{(4)} \ , \nonumber\\
&& \ \ \ P^{(4)}(V_{eff}^{(4)})= \\ &&
\fr{(\sqrt{2}\Delta)^3}{(\sqrt{\pi}g)^3} \int
\delta\left( V_{eff}^{(4)}-
\fr{(\sqrt{2}\Delta)^3}{(\sqrt{\pi}g)^3}
\fr{1}{\eps_2\eps_3}
\right)
\fr{d\eps_2 d\eps_3}{4|\eps_2 \eps_3|\Delta^2}
 \ . \nonumber
\end{eqnarray}
Here the range of variation of $\eps_2$ and $\eps_3$ is the same
as in the Eq.~(\ref{Veff4})
Found in this way the probability distribution in the general
case has the form
\bq\label{Pn}
P^{(n)}(V)\sim \fr{\Delta}{g^{n-1}}\fr{1}{V^2} \chi^{(n)} \ ; \
g^{-n+1}<|V|/\Delta<g^{-1} \ ,
\ee
where $\chi^{(n)}(V)$ is a piecewise polynomial function of the
order $n-3$ of two logarithms $\ln(|V|/\Delta)$ and $\ln g$.
Again in the simplest non-trivial example of $n=4$ one finds
\begin{eqnarray}\label{chi}
\chi^{(4)} =\left\{ \begin{array}{ll}
\ln\left( {|V|g^3}/{\Delta}\right)
\ , \
g^{-3} < |V|/{\Delta}<g^{-2} \\
-\ln\left( {|V|g}/{\Delta}\right)
\ , \
g^{-2} < |V| /\Delta<g^{-1}
\end{array} \right.
\ .
\end{eqnarray}
Also notice the huge range of variation of $V_{eff}^{(n)}$
(\ref{Pn}) for large $n$. Averaging of the matrix element over
the distribution Eq.~(\ref{Pn}) leads to 
\bq\label{Vneff}
\overline{|V_{eff}^{(n)}|}\sim \Delta (\ln
g)^{n-1}/g^n \ .
\ee
Due to a complicated domain of integration over the intermediate
energies it is difficult even to estimate the value of the
overall numerical constant here. However, the important result
of Ref.~\cite{me} is that at large $n$ this unknown overall
constant is neither large like $n!$ nor small like $1/n!$. The
appearance of logarithms in the coefficients of wave function
was also observed in \cite{FF}.

Like in Eq. (\ref{inv}) the corrections generated by the many
particle effective interaction are naturally exponentiated at
least for $n\ll T/\Delta$. Thus combining $\Delta_n$
(\ref{Delt}) with $V_{eff}^{(n)}$ (\ref{Vneff}) we obtain
instead of the Eq. (\ref{inv})
\begin{eqnarray}\label{F}
I &=& \exp \left\{ -\fr{4\sqrt{3}}{\sqrt{\pi}g}
\left(\fr{{\cal E}}{\Delta}\right)^{3/2}
F\left( \fr{\ln g}{g}\fr{{\cal E}}{\Delta}\right)\right\} \ \ , \\
&\,& F(x)=1+\sum F_n x^n \ \ . \nonumber
\end{eqnarray}
In the case of short range screened two-particle interaction
$V(\vec{r}_1,\vec{r}_2) \sim \delta (\vec{r}_1-\vec{r}_2)$ the
first coefficient here is $F_1 = 24\sqrt{2/\pi}/5$. This result
for $I$ may be compared with the perturbative result for the IPR
for disintegration of the single quasi-particle in the isolated
Quantum dot found in Refs.~\cite{AGKL,Mirlin,me}
\bq\label{P5}
I_q =
1 - \fr{1}{\ln g}\phi(y) \ \ , \ \
\phi=\sum_{n=1}^{\infty}\phi_n
y^n \ \ ,
\ee
where $y=(\eps/\Delta)^2\ln g/g$ and $\eps$ is the energy of
quasi-particle. By analogy with the usual IPR, the $I_q$ is a
sum of fourth powers of the amplitudes to find given $k$-th
quasi-particle in the $i$-th exact state $I_q=\sum_i
|\psi_i(k)|^4$.  The next section will be devoted to the more
detailed investigation of the possible properties of the scaling
function $F(x)$. For single quasi-particle the coefficients
$\phi_n$ decrease like $n!^{-1}$ for large $n$, which makes very
improbable the disintegration of quasi-particle through
delocalization transition in the Fock space.

The simple counting of the powers of ${\cal E}/\Delta$ and $g$
in Eq. (\ref{F}) leads to the following scenario of the
development of chaos:

a). For $\Delta g^{2/3} < {\cal E} < \Delta g/\ln g$ the
Eq.~(\ref{inv}) is valid. The number of components in typical
wave function is exponentially large, but described by simple
multiplication of independent direct transitions.

b). At ${\cal E} \sim \Delta g/\ln g$ all terms in the series
Eq.~(\ref{F}) enters the game. The wave function is formed by a
fractal combination of states distant in the Fock space
\cite{Fock}. If the resumed scaling function $F$ will be
a regular function of the argument $x$ (like it is for the
$\phi(y)$ (\ref{P5})), the number of components of the wave
function will be small compared to the total density of states
and the level statistics will be close to Poissonian.

c). Starting from ${\cal E} \sim \Delta g$ all quasi-particles
constituting the wave function have enough energy to decay via
usual Golden-rule.  Only here the Wigner-Dyson statistics is
evident. The more formal criteria for onset of chaos will be the
requirement that the product $I {\cal E} \nu_{total}$ became of
the order of $\sim 1$, where $\nu_{total}$ is the total density
of states described above the Eq.~(\ref{FD}). This condition is
equivalent to the equation for unknown function $F(x)$
\bq\label{cond}
xF(x) = \fr{\pi\sqrt{\pi}}{6\sqrt{2}} \ln g \ .
\ee
Solution of this equation gives the energy of transition to
rigid spectrum somewhere within $g/\ln g < {\cal E}/\Delta < g$.
Due to the exponential character of $I$ and $\nu_{total}$ most
likely this will be the narrow phase transition.

\section{The $n!$ scenario}\label{sec:4}

Thus the power counting for the perturbative IPR Eq.~(\ref{F})
may provide us with the formally self-consistent scenario of
developing of chaos in finite Fermi systems. Now we are going to
show, how taking into account of the high order behavior of the
coefficients $F_n$ may lead to the important revision of this
straightforward scheme. Unfortunately, we are not able to
perform the complete estimation of the asymptotics of $F_n$.
Nevertheless, the catastrophic change of the mechanism of onset
of chaos, which we will consider, may partly compensate the lack
of rigor.  Therefore, even though based on certain assumptions,
the result of this section should be considered as one of the
main results of the paper.

The calculation of any term of the series in Eq.~(\ref{F})
consists of two main steps. First is the calculation of the
density of final states, which we were able to perform
explicitly (\ref{Delt}). The second step is the estimation of
the effective matrix element. The matrix element for a given
Feynman diagram is described by the Eq.~(\ref{Vneff}). As we
have told, this matrix element could not contain any $n!$ or
$n!^{-1}$. However, for many-body problem and high order of
perturbation theory the same initial and final configurations
may be connected by many different Feynman diagrams.  We have
shown on the Fig.~2 the example of corresponding tree-type
$n$-particle diagram. The total number of diagrams like on
Fig.~2 is
\bq\label{ndiag}
\# diagrams = N_d = n^{n-2}(n-1)! n! \ \ .
\ee
Here $n!$ is the number of permutations of $1'-n'$ final
particles, $(n-1)!$ is the number of replacements of (screened)
Coulomb interaction lines and $n^{n-2}$ is the number of
$(n-1)$-segment trees connecting $n$ points $1-n$ \cite{Cayley}.
Eq. (\ref{ndiag}) gives us the number of diagrams of
Schr\"odinger perturbation theory. Among the $N_d$ trees
there are many diagrams having the same set (up to permutations)
of the matrix elements of two-particle interaction $V_{ij}$ and
different energy denominators. As it was pointed out in
Ref.~\cite{me}, these diagrams could not be considered as
statistically independent, which results in lowering of the
mixing by the high order effective interaction. The averaged
modulus of the effective matrix element found for single Feynman
diagram (\ref{Vneff}) should be modified in the case of many
interfering diagrams. However, due to a huge range of variation
of the $V_{eff}$ in the power law distribution Eq.~(\ref{Pn})
the average value is saturated by a rare very large fluctuations
of the matrix element. Therefore for $n$ not too large the
generalization of the Eq.~(\ref{Vneff}) for many diagram case is
achieved by the simple multiplication by the number of diagrams
(we will return later to the discussion of very large $n$).
Finally, combining the Eqs. (\ref{Delt}), (\ref{Vneff}) and
(\ref{ndiag}) we find
\bq\label{Fn}
F_n \le n! \ .
\ee

We see, that the series in Eq. (\ref{F}) most likely is the
asymptotic series. Due to the unsatisfactory estimate of the
number of diagrams and even the value of individual diagram we
were able to draw only the upper bound for the $F_n$. However,
it seems very unlikely that the cancellation between the
diagrams of the Schr\"odinger perturbation theory will remove
the factorial divergency of the coefficients of the series. In
the following (if the opposite not indicated explicitly) we
suppose that $F_n \sim n!$. The asymptotic series, common in the
Field theory and High Energy physics, usually do not cause
serious troubles (see for review Ref.~\cite{Guillou}). In all
known cases one simply breaks the summation on the smallest term
of the series and use this smallest term also as the order of
magnitude estimate of the rest nonperturbative part of the sum.
The more refined strategy will be to perform the Borel
summation, but this procedure (up to the same nonperturbative
corrections) is equivalent to the breaking of the series at the
smallest term.  In our example this means that one should break
the summation in the Eq.  (\ref{F}) at $n_0 \sim g\Delta/\ln g
{\cal E}$ and the non-perturbative ambiguity in $F$ should be
\bq\label{ambig}
\delta F_{np} \sim e^{-n_0} \sim\exp\left\{ - \sigma
\fr{g}{\ln g}\fr{\Delta}{{\cal E}}\right\} \ ,
\ee
with some $\sigma \sim 1$. However, in the known
field-theoretical examples there exist the clear physical
mechanisms (Instanton, Renormalon) which allows to understand
the origin of non-perturbative corrections and the necessity of
breaking the series at the smallest term. In our many-body
quantum problem we could find {\it no physical motivation for
the appearance of nonperturbative corrections of the kind
(\ref{ambig})} and {\it we see no reasons for breaking the
summation in the Eq.  (\ref{F}) at the smallest term}. Of
course, the Eq. (\ref{F}) is not exact, but corrections to
$F(x)$ are small like $(\ln g)^{-1}$ or $\Delta/{\cal E}$.
Taking into account such corrections will help us, to sum up the
original divergent series.  Nevertheless, in order to make sense
of our perturbative analysis of the onset of chaos in finite
Fermi systems we should find the physically motivated cutoff for
the $n!$ divergent sum.  The natural cutoff for $F_n$, which we
see, is given by the total number of particles which may enter
into the diagram of effective interaction $n_{max} \sim
T/\Delta\sim \sqrt{{\cal E}/{\Delta}}$. If $n_{max}> x^{-1}$ the
summation now goes above the smallest term of the series. The
consequences of the decision to sum up the series above the
smallest term are crucial. In the standard scheme the function
$F(x)$ is described by perturbation theory at small $x$ and
became completely nonperturbative (although it still may be
smooth and finite) at $x\sim 1$. Within the new scenario $F(x)$
blows up at
\bq\label{blow}
F_{n_{max}} x^{n_{max}} \sim \left( n_{max}
{\ln g {\cal E}}/{g \Delta}\right)^{n_{max}} \sim 1
\ ,
\ee
which gives the chaotization threshold ${\cal E}_c \sim \Delta
(g/\ln g)^{2/3}$. This estimate was done for the most strongly
divergent series allowed by the Eq. (\ref{Fn}). For example, if
$F_n \sim (n/2)!$ the same calculation will lead to ${\cal E}_c
\sim \Delta(g/\ln g)^{4/5}$. Therefore, the more accurate
conclusion from the Eq. (\ref{Fn}) is that due to the effective
interaction including all $\sim T/\Delta$ excited particles the
series for IPR Eq.  (\ref{inv}) blows up at some $(g/\ln
g)^{2/3}< {\cal E}_c < g/\ln g$ \cite{comment}.

The two important peculiarities of the emergence of Quantum
Chaos follow from this "factorial" scenario:

I. The direct interaction connect only the states close in the
Fock space \cite{Fock}. Therefore, the Eq. (\ref{inv}) describes
mixing of only a small part of close levels. On the other hand,
the highest order effective interaction entering the Eq.
(\ref{blow}) mixes the huge number (the majority of all
many-particle excitations) of very distant simple states. Thus,
divergency of $F(x)$ due to the $n!$ leads to the transition
from Poisson to rigid (Wigner-Dyson) spectrum at ${\cal E}={\cal
E}_c$.

II. Due to the large power of $x$ in the Eq. (\ref{blow}) the
growth of $F(x)$ with increase of energy takes place at a very
narrow interval $\Delta{\cal E}/ {\cal E}_c \sim 1/n_{max}$.
This means that the change of statistics has a form of phase
transition with relative width $\sim 1/n_{max}$.

Our description of the development of chaos is sufficiently
based on the estimate of $n_{max}$ (\ref{blow}). If one would be
able to find another reasons for breaking the series at a term
parametrically smaller than $n_{max} \sim T/\Delta$ the result
for ${\cal E}_c$ will be also different. For example, the new
cutoff may arose due to a huge number of interfering diagrams
Eq.~(\ref{ndiag}). The joint probability distribution
for the sum of many diagrams with power law individual
distributions described by the Eq.~(\ref{Pn}) has a Lorentzian
form \cite{Mello} with the width
$\gamma$ having the form
\bq\label{Lor}
P(V) \sim \fr{\gamma}{V^2+\gamma^2} \fr{\chi(V)}{\chi(\gamma)} \
; \ \gamma\sim \chi(\gamma)\fr{\Delta N_d}{g^{n-1}} \sim
\Delta \fr{n^{3n}}{g^n} \ .
\ee
Here we have used for the estimate of the number of independent
diagrams the upper bound Eq.~(\ref{ndiag}). This distribution is
valid for $|V|<\Delta/g$. The function $\chi$ was defined in the
Eq.~(\ref{Pn}). The factor $\chi(V)/\chi(\gamma)$ allows to
preserve the asymptotics Eq.~(\ref{Pn}) at $|V| \gg \gamma$. Due
to the logarithmic character of $\chi$ this factor correctly
accounts for the deviation from the pure power law in the
distribution of single matrix elements. We see that after taking
into account of the interference of different diagrams the power
law distribution of the total matrix element becomes strongly
violated at $|V|\sim \gamma$, not at $|V|\sim \Delta g^{-n+1}$
as in the Eq.~(\ref{Pn}). The value of $F_n$ found with the
distribution $P(V)$ (\ref{Lor}) becomes parametrically different
from the estimate Eq.~(\ref{Fn}) only at $n'_{max}\sim g^{1/3}$
(even at $n'_{max}\sim (g/\ln g)^{1/3}$ if one considers
seriously the high powers of logarithms in the Eq.~(\ref{Lor})).
However, this new variant of the cutoff leads to the same
threshold for the onset of chaos as we have found before in
Eq.~(\ref{blow}). This coincidence of the results of two
different approaches to estimation of the $n_{max}$ may be
considered as an indirect support for the transition to
Wigner-Dyson statistics at ${\cal E}_c \sim \Delta (g/\ln
g)^{2/3}$.

At least some of the thresholds shown on the Fig.~1 may lie
above ${\cal E}_c$. The physical meaning of these perturbative
thresholds becomes less clear. Nevertheless, even above ${\cal
E}_c$ one may try to look for effects generated by the direct or
low order effective interaction. The highest order effective
interaction becomes important due to the huge combinatorics Eq.
(\ref{ndiag}), but the corresponding matrix element is in
general very small (up to $V\sim \Delta/g^{n_{max}}$). Thus
this high order interaction may be invisible (or at least
suppressed) at small time in time-dependent problems. For
example, one may consider the time evolution of the wave packet
$\Psi(t)$, starting at $t=0$ from single Slater determinant. The
pure direct interaction in this case gives
\bq\label{SGamma}
|\langle \Psi(0)|\Psi(t)\rangle|^2=e^{-\Gamma t} \ ; \ \Gamma
= 2\sqrt{6} \Delta g^{-2}\left({{\cal E}}/{\Delta}\right)^{3/2}
\ .
\ee
The formal power counting (like we have done below the Eq.
(\ref{F})) gives for the range of validity of this result
$\Delta g^{2/3}\ll {\cal E} \ll\Delta g/\ln g $. However, due to
the logarithmic behavior of all integrals (see e.g. the
Eq.~(\ref{Veff4})) all values of the high order effective matrix
elements are equally important within $g^{-n+1}<
|V|/\Delta<g^{-1}$ (\ref{Pn}). Therefore, the range of validity
of the Eq.~(\ref{SGamma}) may be changed strongly after taking
into account of the contribution from large high order effective
matrix elements.

\section{Discussion}\label{sec:5}

The goal of this paper was to consider, how the chaotic mixing
of noninteracting eigenstates develops with increasing the
excitation energy in finite weakly interacting Fermi systems.
Measured in the units of the single-particle level spacing
$\Delta$ the energy corresponding to the onset of rigid spectrum
may depend on only one parameter $g$ (\ref{g}). Nevertheless,
even with only one parameter we were able to introduce a number
of different energy scales, shown on the Fig.~1, which
correspond to (or may correspond to) the different stages of
development of chaos.

The low lying thresholds on the figure $\eps/\Delta \sim  (g/\ln
g)^{1/2}$ and $\eps/\Delta \sim g^{1/2}$ are associated with the
decay of single quasi-particle. These thresholds were considered
in the Refs.~\cite{Sivan,Blanter,AGKL,Mirlin,me} and their
description was not the aim of this paper. However, the methods
introduced in Refs.~\cite{AGKL,me} for the investigation of
below golden rule decay provides us with the useful tool for the
investigation of onset of Chaos in complex many-particle
excitations also.

Technically our consideration is based on the two simple
equations (\ref{P}) and (\ref{inv}) for the IPR. The two
non-interacting states are strongly mixed by the interaction if
the corresponding energy difference turns out to be of the same
order of magnitude with the matrix element. As a result, all the
effects which we consider are determined by the product of the
averaged modulus of the matrix element and the density of
accessible states.

The emergency of chaos in the majority of complicated states
with many excited particles and holes was the main subject of
this paper. This increase of chaos with the growth of excitation
energy proceeds in a very complicated way, both technically and
even more logically. In the simplest variant, mixing of the
simple (single Slater determinant) states is caused by the
direct interaction.  The mixing by direct interaction opens at
${\cal E}> \Delta g^{2/3}$ \cite{room}. Due to a large number of
excited particles the interaction may proceed through a large
number of independent channels. This results in the
exponentiation of the IPR, as we have shown at the end of the
Sec.~2. Thus even the direct interaction may mix together an
exponentially large number of Slater-determinant configurations.
However, this number is still small compared to the total number
of levels of complicated system and the direct interaction could
not cause the transition from Poisson to Wigner-Dyson statistics
of energy levels.

An alternative to direct interaction is the transition via
effective three-particle, four-particle, etc. interaction
considered in Sec.~3.  The effect of this high-order interaction
is much weaker than that of the direct interaction at low
energies, but they became comparable at ${\cal E} \sim \Delta
g/\ln g$. Above the energy ${\cal E} \sim \Delta g/\ln g$ our
perturbative approach is no more valid and we were forced to
introduce the new unknown scaling function $F(\ln g {\cal E}/
g\Delta)$ into the exponent describing the IPR in the
Eq.~(\ref{F}). However, if one expects that $F(x\sim 1)\sim 1$,
the logarithm of the number of components in the typical wave
function at ${\cal E} \sim \Delta g/\ln g$ will be in $\ln g \gg
1$ times smaller than the logarithm of the total number of
states. This means that although the mechanism of mixing is
changed and the number of components in the wave function became
to grow much faster at ${\cal E} \sim \Delta g/\ln g$, this
threshold does not correspond to the transition
Poisson--Wigner-Dyson.

The Wigner-Dyson statistics becomes evident only at ${\cal E}
\sim \Delta g$. This energy is analogous to the golden rule
threshold $\eps \sim \Delta g^{1/2}$ for the decay of single
quasi-particle. Roughly speaking above the ${\cal E} \sim \Delta
g$ any particle from the strip $\sim T$ around $\eps_F$ may
decay via the golden rule.

Formally, in order to understand, how the change of statistics
proceeds between ${\cal E} \sim \Delta g/\ln g$ and ${\cal E}
\sim \Delta g$ one should analyze the high order behavior of the
coefficients $F_n$ in the expansion of the function $F(x)$
Eq.~(\ref{F}). Surprisingly, this analysis in the Sec.~4 instead
of such understanding led us to the crucial revision of the
considered above straightforward scheme.

First of all, as we have shown in the Sec.~4, the series in the
Eq.~(\ref{F}) most likely is the asymptotic series. Although
(\ref{Fn}) $F_n \sim n!$ is only the upper bound for the $F_n$,
it seems very unlikely that the factorial divergency of the
coefficients may be washed out. Moreover, even if $F_n \sim (\nu
n)!$ with any $0< \nu<1$, the results of the Sec.~4 will be only
slightly modified. The factorial divergency of the series would
not be dangerous if one will sum it up e.g. via some variant of
the Borel method. However, within the Borel prescription one is
immediately faced with the problem of strange nonperturbative
corrections like in the Eq.~(\ref{ambig}) (moreover that $F_n$
are positive by construction and the series is non-Borel
summable). Being unable to find the physical motivation for the
Borel summation, we were looking for other way to make sense of
the divergent series. The coefficient $F_n$ is given by the sum
of $n+1$-st order tree type Feynman diagrams. Therefore, the
natural cutoff for the series is given by the largest diagram
which may be drown with available $\sim T/\Delta$ particles.
Namely $n_{max} \sim T/\Delta$ (one more way to determine
$n_{max}$ considered in the Sec.~4 leads to the same physical
result). Within this scenario the $n_{max}$ and the scaling
parameter $x=\ln g {\cal E}/ g\Delta$ in the Eq.~(\ref{F}) are
two independent variables. At $x >1/n_{max}$ summation in the
Eq.~(\ref{F}) goes above the smallest term.  However, up to the
energy ${\cal E}_c\sim \Delta(g/\ln g)^{2/3}$ all $n_{max}$
terms of the series are small. At ${\cal E}={\cal E}_c$ even the
broken series blows up and above ${\cal E}_c$ our perturbative
approach is no more valid. Because of the growth of $F(x)$ at
${\cal E}\approx {\cal E}_c$ proceeds mostly due to the last
terms of the series or the most complicated diagrams of the
highest allowed order of perturbation theory, this threshold
corresponds to narrow phase transition from Poisson to
Wigner-Dyson spectrum. We have attempted at the end of the
previous section to find remnants of non-chaotic behavior above
${\cal E}_c$ by considering the short time processes. However,
this issue requires a separate and more careful investigation.

To summarize, we have considered in this paper the two
scenarios, how the quantum chaos may develops in the Finite
Fermi systems depending on the way one will use for the
summation of factorially divergent contributions from the high
order effective interaction.
\begin{itemize}
\item If the series is summed up within the Borel prescription
(i.~e. broken effectively at the smallest term) we were able to
find a few distinct stages of the chaotic behavior. Below the
energy $\sim \Delta g^{2/3}$ the majority of eigenvectors is
almost unperturbed. Between $\sim \Delta g^{2/3}$ and $\sim
\Delta g/\ln g$ direct interaction mixes together the
exponentially large number of simple states (\ref{inv}).
However, the IPR is under theoretical control in this region and
the statistics is clearly Poissonian. Above $\Delta g/\ln g$
statistics of the energy levels is still Poissonian, but the IPR
is described by some (unknown) scaling function $F(x)$
(\ref{F}). Transition to Wigner-Dyson takes place somewhere
between ${\cal E} \sim \Delta g/\ln g$ and ${\cal E} \sim
\Delta g$ (see the Eq.~(\ref{cond})). The transition may be
relatively narrow $\Delta{\cal E}/{\cal E} \sim (\ln g)^{-1}$.
\item In the second variant the cutoff of the series in the
Eq.~(\ref{F}) is determined by the number of the particles in
the largest allowed Feynman diagram and is independent of the
scaling variable $x$. In this case the transition to Wigner-Dyson
takes place at ${\cal E}_c\sim \Delta(g/\ln g)^{2/3}$ and is
narrow like $\Delta{\cal E}/{\cal E} \sim \Delta/T \sim (\ln
g/g)^{1/3}$. 
\end{itemize}
To our current understanding, the second scenario is more
physically motivated. However, our argumentation in this part is
not completely rigorous and one still may find the reasons for
summation a la Borel of the tree diagrams. The important
disadvantage (or advantage?) of the second scheme is that if it
is correct, it will be probably the first example of the phase
transition caused by the factorial divergency of the series.
Finally, although both of our scenarios are reach of new
physical predictions, we were able to make only the order of
magnitude estimates of the considered thresholds. Therefore, any
experimental confirmation (within the true experiment, or by
numerical simulations) of the mechanisms of development of chaos
considered in this paper are very desirable.

{\bf Acknowledgments.} Author is thankful for discussions to
V.F.Dmitriev, V.V.Flambaum, A.A.Gribakina, G.F.Gribakin,
I.V.Ponomarev, D.V.Savin, V.V.Sokolov, O.P.Sushkov,
V.B.Telitsin, and A.S.Yelkhovsky. The work has been supported in
part by the Gordon Godfry foundation and by RFBR, grant
98-02-17905.

\begin{figure}[t]
\epsfxsize=8cm
\epsffile{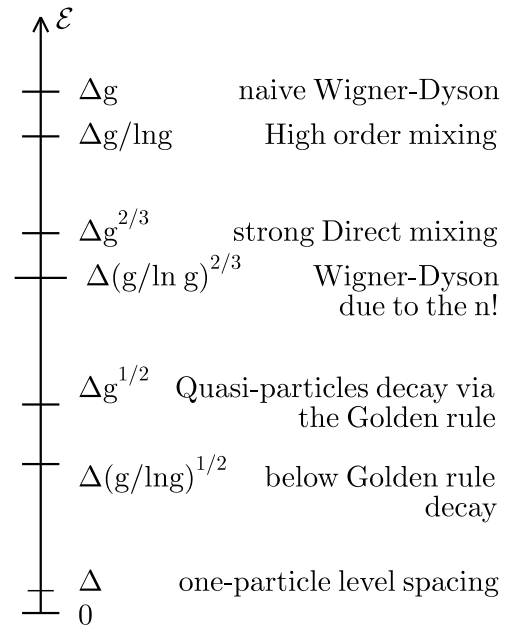}
\vglue 0.2cm
\medskip
\caption{The different energy scales corresponding to
chaotization of single- or many-particle excited states.
}
\end{figure}

\begin{figure}[t]
\epsfxsize=7.5cm
\epsffile{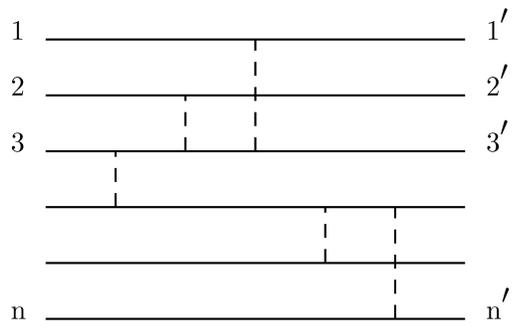}
\vglue 0.2cm
\medskip
\caption{The tree type diagram contributing to high order
effective matrix element.
The screened Coulomb interaction is shown by dashed lines.
}
\end{figure}


\begin{thebibliography}{99}
\vspace{-1cm}

\bibitem{Wigner} E.~P.~Wigner, Ann. Math. {\bf 62}, 548 (1955); {\bf
65}, 203 (1957).

\bibitem{Wong} J.~B.~French and S.~S.~M.~Wong, Phys. Lett.
{\bf 33B}, 447 (1970); Phys. Lett. {\bf 35B}, 5 (1971).


\bibitem{Bohigas} O.~Bohigas and J.~Flores, Phys. Lett. {\bf
34B}, 261 (1971); Phys. Lett. {\bf 35B}, 383 (1971).

\bibitem{Zelev} M.~Horoi, V.~Zelevinsky and B.~A.~Brown,  Phys.
Rev. Lett. {\bf 74}, 5194 (1995);  V.~Zelevinsky, B.~A.~Brown,
N.~Frazier and M.~Horoi, Phys. Rep. {\bf 276}, 85, 1996.

\bibitem{Flamb} V.~V.~Flambaum, G.~F.~Gribakin and
F.~M.~Izrailev, Phys. Rev. E {\bf 53}, 5729 (1996).

\bibitem{Pichard} D.~Weinmann, J.-L.~Pichard and Y.~Imry,
cond-mat/9705113.

\bibitem{Shepelyansky} Ph.~Jacqoud and D.~L.~Shepelyansky,
cond-mat/9706040; B.~Georgeot and D.~L.~Shepelyansky,
cond-mat/9707231.

\bibitem{Weidenmuller} T.~Guhr, A.~M\"uller-Groeling and
H.~A.~Weidenm\"uller, cond-mat/9707301, Phys. Reports (to
appear)

\bibitem{SImry} U.~Sivan, F.~P.~Milliken, K.~Milkove,
S.~Rihston, Y.~Lee, J.~M.~Hong, V.~Boegli, D.~Kern and
M.~deFranza,  Europhys.
Lett. {\bf 25}, 605 (1994).

\bibitem{Sivan} U.~Sivan, Y.~Imry and A.~G.~Aronov, Europhys.
Lett. {\bf 28}, 115 (1994).

\bibitem{Blanter} Ya.~M.~Blanter, Phys. Rev. {\bf B 54}, 12807
(1996).

\bibitem{AGKL} B.~L.~Altshuler, Y.~Gefen, A.~Kamenev and
L.~S.~Levitov, Phys. Rev. Lett. {\bf 78}, 2803 (1997).

\bibitem{Mirlin} A.~D.~Mirlin and Y.~V.~Fyodorov, Phys. Rev.
{\bf B 56}, 13393 (1997).

\bibitem{me} P. G. Silvestrov, Phys. Rev. Lett. {\bf 79}, 3994
(1997).

\bibitem{endnote} We suppose that the Fermi energy is large
compared to the excitation energy $\eps_F \gg {\cal E}$. The
number of particles goes to infinity but experimental accuracy
still allows to resolve individual energy levels (the mesoscopic
limit).

\bibitem{Bohr} A.~Bohr and B.~R.~Mottelson, {\it Nuclear Structure},
Benjamin, New York {\bf 1}, 284 (1969).

\bibitem{BlBl} The temperature and total energy may be easily
related via ${\cal E}=\int (\eps-\eps_F) \overline{n(\eps)}
d\eps/\Delta$.

\bibitem{Ponomarev} I.~V.~Ponomarev, P.~G.~Silvestrov, Phys.
Rev. {\bf B 56}, 3742 (1997).

\bibitem{FF}V.~V.~Flambaum, F.~M.~Izrailev, Phys. Rev. {\bf E
56}, 5144 (1997).

\bibitem{Fock} We use the concept of the distance in the Fock
space introduced in Ref. \cite{AGKL}. The distance between two
simple states equals to one if they may be related by the direct
interaction in the first order of perturbation theory.

\bibitem{Cayley} This combinatorial calculation was done by Cayley
more then century ago. Author is thankful to A.~Gribakina and
G.~Gribakin for discussion of this result.

\bibitem{Guillou} {\it Large order behavior of perturbation
theory}, edited by L.~C.~Le~Guillou and J.~Zinn-Justin
(North-Holland, Amsterdam, 1990).

\bibitem{comment} Only if the factorial divergency of $F_n$ will
be washed out completely due to the large number of coinciding
diagrams in Eq. (\ref{ndiag}) (which seems very unlikely)
${\cal E}_c$ will be raised again to that predicted by the
Eq.~(\ref{cond}).

\bibitem{Mello} See for example the Appendix B in the lecture by
P.~A.~Mello in:{\it Mesoscopic Quantum Physics} Les Houches,
Session LXI 1994 (Elsevier, Amsterdam, 1995)

\bibitem{room} In particular this means that the decay of single
quasi-particles starts at parametrically lower energies than the
mixing of complicated many-particle excitations ($\Delta
g^{1/2}\ll \Delta g^{2/3}$). Thus below the energy $\eps \sim
\Delta g^{2/3}$ one may consider the decay of quasi-particle (via
the golden rule, or not) into almost nonperturbed few particles
states as it was done in
Refs.~\cite{Sivan,Blanter,AGKL,Mirlin,me}.

\end{thebibliography}
\end{document}